\documentclass[english,aps]{iopart}
\usepackage[T1]{fontenc}
\usepackage[latin9]{inputenc}
\usepackage{graphicx}
\usepackage{amssymb}

\makeatletter
\usepackage{iopams}
\usepackage{setstack}

\makeatother

\usepackage{babel}

\begin{document}

\title{Spectral measurement of the thermal excitation of a superconducting
qubit}

\author{A. Palacios-Laloy, F. Mallet, F. Nguyen, F. Ong, P. Bertet, D. Vion,
D. Esteve}
\begin{abstract}
We report the measurement of the fluctuations of a transmon qubit
through the noise spectrum of the microwave signal that measures its
state. The amplitude of the Lorentzian noise power spectrum allows
to determine the average qubit excitation, in agreement with the estimated
thermal radiation reaching the sample. Its width yields the qubit
energy relaxation rate which decreases with temperature, contrary
to the predictions for a two-level system solely coupled to thermal
radiation. This indicates the existence of another non-radiative energy
relaxation channel for the qubit.
\end{abstract}
\maketitle
Superconducting qubits \cite{clarke_superconducting_2008} are promising
candidates for implementing a solid-state quantum processor. Over
the last years, substantial improvements have been made in the coherence
times \cite{vion_quantronium_2002,schreier_suppressing_2008}, fidelity
of single-qubit gates \cite{chow_randomized_2009}, readout procedures
\cite{mallet_2009,mcdermott_simultaneous_2005} and entanglement of
several qubits \cite{steffen_entanglement_2006}. In a recent experiment
a simple quantum algorithm was operated on a two-qubit elementary
processor \cite{dicarlo_demonstration_2009}. One of the requirements
for the implementation of larger scale quantum algorithms \cite{diVincenzocriteria}
is that the qubit registers should be properly initialized at the
beginning of each computation, with all qubits lying in their ground
state. In most superconducting qubit experiments, the initialization
is simply realized by waiting long enough before each experimental
sequence for the system to reach thermal equilibrium. At cryogenic
temperatures in the $10-30\,$mK range and for typical qubit resonance
frequencies of a few GHz, there is indeed at thermal equilibrium a
small (typically less than $1\%$) probability of finding the qubits
in the excited state, which is usually considered negligible. However,
given the recent improvement of the overall fidelity of single- and
two-qubit gates, the effect of even small thermal fluctuations will
require to be considered more quantitatively in the near future. Moreover
it is well known in mesoscopic physics that the effective temperature
of an electrical degree of freedom such as a superconducting qubit
can be in some cases much larger than the temperature of the cryostat,
because it can be strongly coupled to out-of-equilibrium electromagnetic
radiation coming from the measuring leads, while only weakly to the
phonon bath. It is thus important to be able to measure precisely
the average excited state population of a single qubit.

Here we propose and demonstrate a method to determine this thermal
excited state population in a circuit Quantum Electrodynamics (cQED)
setup \cite{blais_cavity_2004}, where a Cooper-pair box qubit of
the transmon type \cite{koch_charge-insensitive_2007,schreier_suppressing_2008}
is coupled to a coplanar waveguide resonator (CPW). The two qubit
states shift differently the resonator frequency, so that the phase
of a microwave signal reflected by the resonator allows a non-destructive
readout as demonstrated in numerous experiments \cite{wallraff_strong_2004}.
In the present setup as in most cQED experiments, it is not possible
to readout the qubit state in one single experimental sequence due
to insufficient signal-to-noise ratio. Note however that such a single-shot
readout was recently obtained in cQED by using a non-linear CPW resonator
\cite{mallet_2009}. The usual method for reading out the qubit state,
using ensemble-averaged measurements of the microwave signal, does
not directly provide an absolute measurement of the qubit excitation.
However, thermal fluctuations of the qubit state are responsible for
a measurable phase noise in the microwave signal reflected by the
resonator, with a characteristic Lorentzian power spectrum. In this
work we report the observation of this thermal noise and we use it
to determine the effective qubit temperature. We note that a related
measurement was performed on an ensemble of nuclear spins measured
by a SQUID amplifier\cite{PhysRevLett.55.1742}.

\begin{figure}
\begin{centering}
\includegraphics[width=7cm]{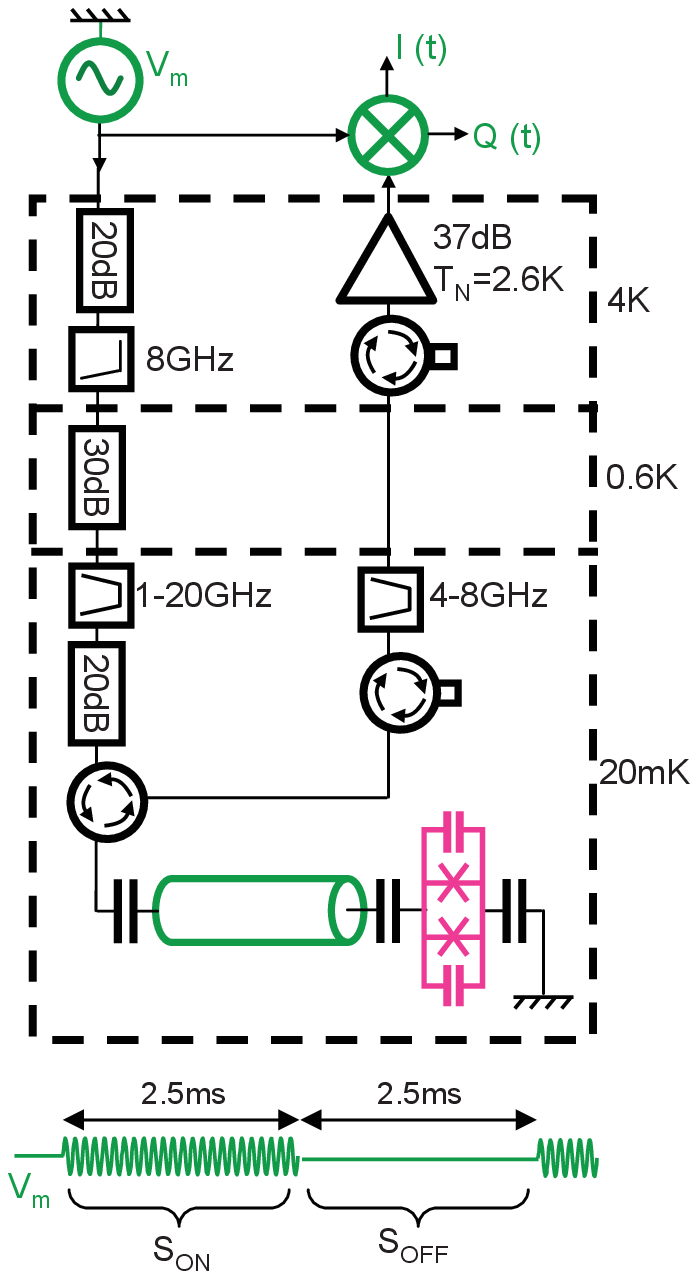}
\par\end{centering}

\caption{\label{fig:setup}Microwave setup used for measurements. A microwave
signal $V_{m}$ is sent through the input line containing several
attenuators and filters at each temperature stage to the input port
of the resonator (shown in green). The reflected signal is separated
from the input one by a circulator and goes through a filter and two
isolators before reaching the cryogenic amplifier (gain $37\,$dB
and noise temperature $2.6\,$K). The signal is then demodulated at
room temperature with a homodyne demodulation scheme to get its in-phase
and quadrature components $I(t)$ and $Q(t)$ respectively (for sake
of simplicity details of the room temperature demodulation scheme,
including several stages of amplification are not shown). Switching
the signal ON and OFF with 1ms period allows to substract the noise
background coming from the amplifier.}

\end{figure}
The complete experimental setup is shown in Fig \ref{fig:setup}.
The transmon has its two lowest energy eigenstates $\left|g\right\rangle $
and $\left|e\right\rangle $ separated by $\omega_{ge}/2\pi=$5.304$\,$GHz.
It is capacitively coupled with strength $g/2\pi\,$=$\,$45$\pm2\,$MHz
to a superconducting resonator with resonance frequency $\omega_{c}/2\pi=\,$5.796$\,$GHz
and bandwidth $BW=\,$30.3$\,$MHz, which serves as qubit readout.
With these parameters, the qubit is sufficiently detuned from the
resonator for their interaction to be well described by the dispersive
Hamiltonian $H=\hbar\chi\hat{n}\hat{\sigma}_{z}$, where $\chi$ is
the dispersive coupling constant and $\hat{n}$ is the intra-resonator
photon number operator. The resonator frequency is thus shifted by
$\pm\chi/2\pi=1.75\,$MHz when the qubit is in $\left|g\right\rangle $
or $\left|e\right\rangle $ respectively. A continuous microwave tone
of frequency $\omega_{c}/2\pi$ sent to the resonator input from source
$V_{m}$ acquires a qubit state-dependent phase shift which allows
a continuous and non-destructive monitoring of this state. This continuous
measurement does not induce spurious qubit excitation as long as the
the intra-resonator photon number ($\overline{n}\simeq2.5$ in our
measurements) is much below the critical photon number $n_{crit}=\left(\omega_{ge}-\omega_{c}\right)^{2}/4g^{2}\simeq30$
above which the dispersive approximation fails. After reflection on
the resonator, the signal is routed through a circulator to a cryogenic
amplifier and is then measured by homodyne detection at room-temperature,
yielding the two field quadratures $I(t)$ and $Q(t)$. The thermal
fluctuations of the qubit state induce some phase noise on the reflected
microwave signal, and thus some noise on each quadrature $X(t)$ ($X=I,Q$). 

We start by computing the power spectrum of the qubit thermal fluctuations.
Assuming that the bath consists of a bosonic Markovian bath at temperature
$T$, as expected for the impedance of the electromagnetic environment,
the qubit dynamics at thermal equilibrium is described by a simple
rate equation \cite{atomphoton}

\begin{equation}
\dot{\rho_{ee}}=-\dot{\rho_{gg}}=-\Gamma\rho_{ee}+\Gamma n_{th}(1-2\rho_{ee})\label{eq:dynamics}\end{equation}

\noindent where $\Gamma$ is the qubit energy relaxation rate, and
$n_{th}(T)=(\exp(\hbar\omega_{c}/kT)-1)^{-1}$ is the mean photon
number at temperature $T$. This yields a steady-state population
of the qubit excited state $\rho_{ee}^{th}=\frac{n_{th}}{1+2n_{th}}$,
or $z_{th}=-\frac{1}{1+2n_{th}}$ after conversion into spin units
$z(t)=2\rho_{ee}(t)-1$. The corresponding noise power spectrum $S_{z}(\omega)$
can be computed as the Fourier Transform of the two-time correlation
function $C_{z}(\tau)=\left\langle z(\tau)z(0)\right\rangle $ which
is

\begin{equation}
C_{z}(\tau)=4\:\exp\left(-\Gamma(1+2n_{th})\tau\right)\left[1-\rho_{ee}^{th}\right]\,\rho_{ee}^{th}.\end{equation}

\noindent After Fourier transform, we obtain 

\begin{equation}
S_{z}(\omega)=4\,\frac{\Gamma(1+2n_{th})}{\Gamma^{2}(1+2n_{th})^{2}+\omega^{2}}\left[1-\rho_{ee}^{th}\right]\,\rho_{ee}^{th}.\label{eq:sz}\end{equation}

Note that these expressions are only approximate because the transmon
is not a genuine two-level system but an anharmonic resonator with
an infinite number of excited states. The previous expressions are
thus only valid in the limit where the population of these higher
excited states is negligible, which in our case is true up to temperatures
around $100\,$mK. 

We model the effect of the qubit state thermal fluctuations on the
measuring signal quadratures $X(t)$ by assuming that the field inside
the resonator follows instantaneously the qubit state. Here this assumption
is justified by the large bandwidth of the resonator, obtained by
chosing a large resonator input capacitor. The quadratures are then
simply expressed as $X(t)=\overline{X}+(\Delta X/2)z(t)+\xi(t)$,
where $\overline{X}$ is the average reflected signal for a qubit
fully unpolarized, $\Delta X$ is the change in $X$ when the qubit
changes state, and $\xi(t)$ is the total output noise of the amplifier.
In the experiment we measure the sum of the noises on both quadratures\begin{equation}
S_{V,ON}(\omega)=S_{I}(\omega)+S_{Q}(\omega)=S_{\xi}(\omega)+(\Delta V/2)^{2}S_{z}(\omega),\label{eq:svon}\end{equation}

\noindent where $S_{\xi}(\omega)$ is the output amplifier noise power
spectrum and $(\Delta V/2)^{2}=(\Delta I/2)^{2}+(\Delta Q/2)^{2}$
is the detector sensitivity. This quantity has the advantage of being
insensitive to drifts of the phase between the local oscillator used
in the demodulation and the measurement signal. It is worth noting
that the mere presence of a continuous measurement of the qubit state
has no effect on the dynamics of thermal fluctuations because this
dynamics is fully incoherent and entirely goverened by Markovian rate
equations \cite{bernu_zeno_2008}. The situation is very different
when the qubit is continuously measured while being coherently driven,
in which case the dynamics changes from diffusive Rabi oscillations
to quantum jumps when the measurement strength is increased \cite{palacios_2009}.

We measure the detector output noise spectrum $S_{V}(\omega)$ for
a series of temperatures $T_{c}$. Each spectrum is measured after
waiting $15$ minutes thermalization time once the cryostat reaches
$T_{c}$. We also verify that the sample is well thermalized by acquiring
two noise spectra for each $T_{c}$, one upon warming up and the second
upon cooling down, which are found to be nearly identical. Each spectrum
is acquired by sampling $I(t)$ and $Q(t)$ with $100\:$MHz sampling
frequency. Each 1024-point set of the sampled signals is Fourier transformed
and the amplitude of this transform is squared to obtain the noise
spectra $S_{I}(\omega)$ and $S_{Q}(\omega)$, which are then corrected
for the variations of setup gains in frequency and between the $I$
and $Q$ channels. The resulting spectra are summed to form $S_{V,ON}(\omega)$.
Each $2.5\:$ms the measurement microwave is turned OFF to measure
the noise background of the amplifier $S_{V,OFF}(\omega)=S_{\xi}(\omega)$
and subtract it from the signal. The resulting noise spectrum $S_{V}(\omega)=S_{V,ON}(\omega)-S_{V,OFF}(\omega)$
is averaged typically $10^{6}$ times.

\begin{figure}
\begin{centering}
\includegraphics[width=12cm]{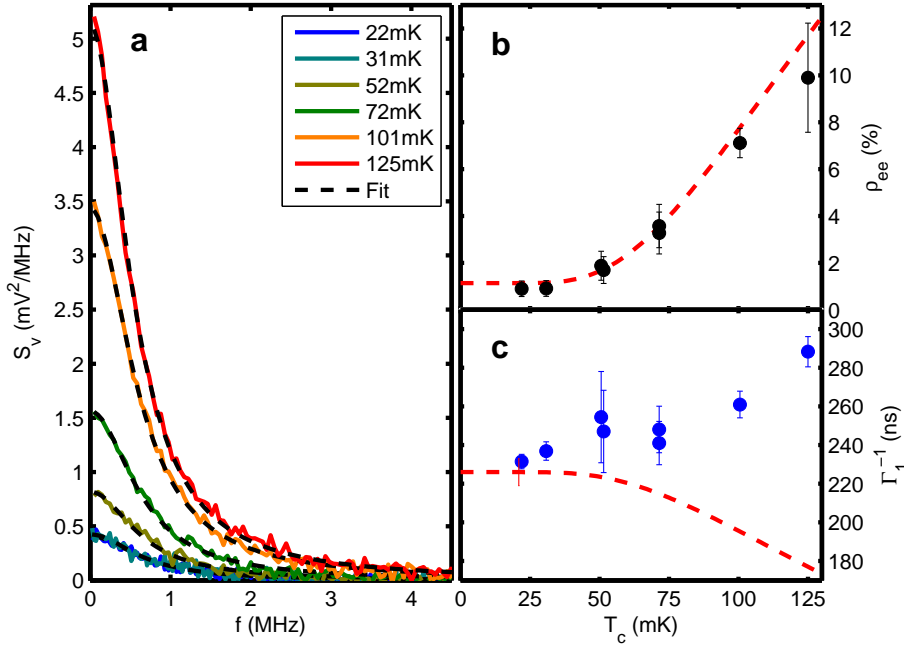}
\par\end{centering}

\caption{\label{fig:(a)-Noise-spectra}(a) Noise spectra acquired for several
temperatures $T_{c}$ (color solid lines) and lorentzian fits (dashed
black lines). (b) Thermal population of the TLS as a function of temperature:
comparison of the experimental data (black dots) with the theoretical
prediction (red dashed curve). (c) Relaxation times as a function
of temperature extracted from the widths of the loretzian spectra
(blue dots) compared to the predictions of the model discussed in
the text (red dashed curve) taking $\Gamma^{-1}=T_{1,20mK}=226\pm7\,$ns,
independently measured in a pulsed experiment at $20\:$mK.}

\end{figure}
As shown in Fig 2a, the measured noise spectra $S_{V}(\omega)$ have
a Lorentzian shape, with an amplitude rapidly increasing with temperature.
The amplitude $A$ and width $\Gamma_{1}$ of each spectrum are fitted
with a Lorentzian model $A\,\Gamma_{1}/(\Gamma_{1}^{2}+\omega^{2})$.
According to Eqs \ref{eq:sz} and \ref{eq:svon}, the model predicts
that $\Gamma_{1}=\Gamma(1+2n_{th})$ and $A=\Delta V^{2}\left[1-\rho_{ee}^{th}\right]\,\rho_{ee}^{th}$.
The detector sensitivity $\Delta V^{2}$ is experimentally calibrated
in the following way : using exactly the same setup, we ensemble-average
$V^{2}(t)=I^{2}(t)+Q^{2}(t)$ under saturation of the qubit $g\rightarrow e$
transition with a second microwave source at frequency $\omega_{ge}/2\pi$.
This yields $\Delta V/2=2.76\pm0.14\,$mV. In this way we can directly
extract from the fits the thermally excited state population $\rho_{ee}$
and the relaxation rates $\Gamma_{1}$ as a function of the cryostat
temperature $T_{c}$. 

The fitted population (dots in Fig. \ref{fig:(a)-Noise-spectra}b)
agrees with the theoretical average population $\rho_{ee}^{th}$ (red
dashed curve), calculated assuming two sources of radiation : the
thermal field corresponding to the temperature of the cryostat coldest
stage $T_{c}$, with an average of $n_{th}(T_{c})$ photons, and the
thermal field radiated by the $30\:$dB attenuator thermalized at
the still temperature $T_{S}=$600$\pm$100 mK, and attenuated ($22\pm0.5\,$dB)
at $20\:$mK, contributing with $n_{th}\left(T_{S}\right)/10^{2.2}$
photons. At the lowest $T_{c}$, we find a thermally excited state
population of $1\pm0.5\%$, corresponding to an effective temperature
of $55\:$mK. 

At $T_{c}=20\:$mK, the relaxation rate $\Gamma_{1}^{-1}$ deduced
from the width of the Lorentzian noise spectrum (see Fig. \ref{fig:(a)-Noise-spectra}c)
is found to be in excellent agreement with the qubit relaxation time
$T_{1,20mK}=226\pm7\:$ns, measured in a standard pulsed sequence.
However, at higher $T_{c}$, we observe that the fitted width decreases,
which implies that the qubit energy relaxation time increases with
temperature. This couterintuitive result disagrees with our model
which predicts a relaxation rate $\Gamma(n_{th})=\Gamma(1+2n_{th})$
(see Eq. \ref{eq:dynamics}) increasing with temperature due to stimulated
emission by the thermal field, yielding the red dashed curve in Fig.
\ref{fig:(a)-Noise-spectra}b (calculated with $\Gamma=T_{1,20mK}^{-1}$).
This indicates that the qubit is not only coupled to its electromagnetic
environment but also to another type of bath, causing some additional
damping with a different temperature dependence. Additional support
for this idea is that the measured relaxation time at $20\:$mK ($226\:$ns)
is significantly shorter than the expected damping time due to relaxation
into the the external impedance at zero temperature ($600\:$ns),
which indicates the existence of a non-radiative energy decay channel.
We finally note that a similar increase of the relaxation time with
temperature up to $150\:$mK was directly observed in a superconducting
phase qubit, and attributed to non-equilibrium quasiparticles in the
superconducting metal electrodes \cite{martinis_energy_2009}; a similar
scenario might explain our results. 

In conclusion, we have determined the thermal population of a superconducting
qubit coupled to a resonator, even without single-shot detection capability,
by studying the noise spectrum of its measuring signal. The population
measured is in good agreement with the estimated thermal radiation
reaching the sample. We observe however an increase in the relaxation
time with temperature in contradiction with this model. This points
to the existence of unknown non-radiative decay channels as observed
in other qubit experiments \cite{martinis_energy_2009}.

\ack{}{We acknowledge financial support from European project EuroSQIP,
Agence Nationale de la Recherche (grant ANR-08-BLAN-0074-01), and
Region Ile-de-France for the nanofabrication facility at SPEC. We
gratefully thank P. Senat, P. Orfila and J.-C. Tack for technical
support, and acknowledge useful discussions within the Quantronics
group.}

\bigskip{}

Correspondance should be addressed to P.B.

\bibliographystyle{iopart-num}
\addcontentsline{toc}{section}{\refname}\bibliography{thermal-refs}

\end{document}